# Information Equation of State

Paul Gough,
Space Science Centre,
University of Sussex,
Brighton, BN1 9QT, UK.
Email: m.p.gough@sussex.ac.uk

**Abstract**
Landauer's principle is applied to information in the universe. Once stars began forming, the increasing proportion of matter at high stellar temperatures compensated for the expanding universe to provide a near constant information energy density. The information equation of state was close to the dark energy value, $w_i = -1$, for a wide range of redshifts, $10 > z > 0.8$, over one half of cosmic time. A reasonable universe information bit content of only $10^{87}$ bits is sufficient for information energy to account for all dark energy. A time varying equation of state with a direct link between dark energy and matter, and linked to star formation in particular, is clearly relevant to the cosmic coincidence problem. In answering the 'Why now?' question we wonder 'What next?' as we expect the information equation of state to tend towards $w = 0$ in the future.

**Keywords:** dark energy theory, cosmological constant experiments, classical tests of cosmology

## 1. Introduction.

Landauer showed that any erasure of information is necessarily accompanied by heat dissipation [1]. When a bit in the memory of a computer is forced to a standard state, usually by being reset to zero, irrespective of its present value, which may not be known, the information bearing degrees of freedom and thus the entropy of the computer memory are clearly reduced. A corresponding minimum $k\,T\,ln2$ of heat energy has to be dissipated into the surrounding environment to increase the environment's entropy in compensation, and in accord with the second law of thermodynamics. The total amount of information is conserved as the surrounding environment effectively contains the erased information, although clearly no longer in a form that the computer can use.

Landauer's principle was originally proposed to describe the energy dissipation when information is overwritten in computer systems and subsequently used to predict the future limits to shrinking computer circuit size [2]-[9]. Shannon information entropy of a computer memory device, $log_2 N_L$, is just the number of bits needed to distinguish between the $N_L$ information bearing degrees of freedom used for computer logic states [10]. In comparison, thermodynamic entropy, $k\,lnN$ accounts for the $N$ discrete Boltzmann microscopic states of all constituent atoms, electrons, etc. In equivalent units, the thermodynamic entropy of such a computer memory device is therefore many orders of magnitude larger than the information entropy. The energy dissipated by present day computer memory through the Landauer's principle is thus miniscule in comparison to normal thermal dissipation. However, information entropy is equivalent to thermodynamic entropy when the same degrees of freedom are considered. The information entropy of the physical world is thus the number of bits needed to account for all possible microscopic states. Then each bit of information is equivalent to $\Delta S = k_B\, ln2$ of thermodynamic entropy leading to Landauer's principle that $\Delta S\, T = k_B\, T\, ln2$ of heat is dissipated when a computer logic bit is erased.

Information is therefore directly bound up with the fundamental physics of nature. This strong interdependence between nature and information is emphasized by astrophysicist John Wheeler's slogan "it from bit" and computer scientist Rolf Landauer's maxim "information is physical".

## 2. Applying Landauer's principle to the universe.

In the computer memory example of our introduction information 'erasure' was shown to be a misnomer as the information present in the computer logic is not destroyed on memory reset but just converted into a form that computer logic can no longer access. Henceforth we enclose the term 'erasure' in quotes to represent this action. Each bit of logic 'erased' is converted into $k_B\, T\, ln2$ of heat energy with an associated increase in the states of the environment surrounding the memory device and overall information is conserved. More generally, Landauer's principle applies to all systems in nature so that any system, temperature T, in which information is 'erased' by some physical process will output $k_B\, T\, ln2$ of heat energy per bit 'erased' with a corresponding increase in the information of the environment surrounding that system.

Landauer's principle is fully compatible with the laws of thermodynamics [9], [11-14]. Indeed Landauer's principle has been used to reconcile the operation of Maxwell's Demon with the second law of thermodynamics. The 'erasure' of information when the Demon resets its one bit of memory in preparation for the next measurement is the source of heat dissipation that ensures that there is no overall reduction in entropy [3][9][11]. Landauer's principle can be derived from microscopic considerations [13] as well as derived from the well-established properties of the Shannon-Gibbs-Boltzmann entropy [14]. The principle thus appears to be fundamental and universal in application.

Landauer's principle applies both to classical and to quantum information. A foundational principle of quantum mechanics has been proposed [15] that all elementary systems carry one bit of information. When N initially separate and independent elementary systems interact the assemblage will still carry N bits of information, even if there is complete entanglement, provided there is no information exchange with the environment [15]. Recently Landauer's principle has been applied to quantum information theories of vacuum entanglement, holographic dark energy, and black hole physics [16-17]. For example, dark energy may arise from quantum information loss at the cosmic horizon [17].

Information in the universe is continuously operated on and changed by physical processes [18]. It is estimated that the universe contains some $10^{90}$ bits and that around $10^{120}$ bits have been processed to present [18-20]. Note that $10^{90}$ bits is significantly less than the theoretical maximum universe information content of $10^{123}$ bits provided by applying the holographic principle [18][21] to the universe's surface area. In the maximum information scenario, corresponding to the universe being one single black hole, $10^{123}$ elemental squares of Planck length size are needed to cover the surface of the present known universe.

In the regular operation of computers there are many occurrences of data overwriting and resetting with associated heat dissipation from information 'erasure' the norm and only avoided by utilising special reversible logic circuits or procedures [1][2][4-5][8]. Thus it seems reasonable to assume that some natural physical processes may also lead to similar heat dissipation from 'erasure' of some of the $10^{90}$ bits of information intrinsic to the physical world. Fortunately, for the purpose of deriving this information energy contribution to the universe energy balance it is not necessary to quantify what fraction of the information processed is being, or has been, 'erased' in these processes, or even to identify the possible physical processes responsible for 'erasure'. It is sufficient to be able to say information may be 'erased' and thus allow us to equate each bit to $k_B \, T \, log2$ of energy that can be dissipated as heat. Such equivalence is standard in cosmology and not fundamentally different from the way we use $mc^2$ to represent the contribution of all mass to the total universe energy budget when to date only a small fraction of baryon mass has been converted into fusion energy within stars. Most importantly note the equivalent energy of a bit depends solely on the temperature of the system of microstates described.

This work takes an information based approach concentrating on just the recent epoch when stars were formed, in contrast to other works which consider the more difficult accounting of information entropy over the whole time span since the big bang [18-20][22-23].

## 3. Information equation of state: just prior to star formation.

The universe temperature for the period of time between decoupling and the commencement of any significant star formation is given approximately by the temperature, $T$, of a radiating universe with the same energy density as our mass dominated universe, with total mass density, $\rho_{tot}$ (including dark matter):

$$\rho_{tot} \, c^2 = \sigma_v \, T^4 \qquad (1)$$

The (volume) radiation constant, $\sigma_v$, is defined in terms of the Stefan-Boltzmann (area) constant, $\sigma_a$, by $\sigma_v = 4\sigma_a/c$. Replacing $\sigma_a$ by its definition in terms of fundamental constants and applying Landauer's principle we obtain equation (2), the energy, $E$, associated with each bit:

$$E = k_B T \, ln2 = (15 \, \rho_{tot} \, \hbar^3 \, c^5 / \pi^2)^{1/4} \, ln2 \qquad (2)$$

Equation (2) shows that the characteristic bit energy, $E$, was proportional to $\rho^{1/4}$, and thus to $a^{-3/4}$, where $a$ is the universe scale size. We expect from the second law that total information did not decrease and so we can assume that the evolution of the universe total information bit content lay between two limits. At one limit (most likely) the total bit content remained fixed with the information bit number density falling in parallel with the density of matter as $a^{-3}$. At the other limit the number of microstates increased with increasing volume to provide a constant bit number density. These simply defined limits have been chosen to be broad enough to include previously considered cases [18-20][22-23] and at the same time still provide a clear comparison against the period of star formation discussed next. Thus information energy density varied with universe scale size as $a^{-3.75}$ if total information bit content remained constant, and as $a^{-0.75}$ if information bit density was constant. These variations (as $a^{-3(1+w)}$ where $w$ is the equation of state) correspond to an information equation of state, $w_i$, just prior to star formation bounded by the range: $-0.75 < w_i < +0.25$.

## 4. Information equation of state: during star formation

Landauer's principle clearly identifies temperature as the single parameter relating information to energy. As stars formed an increasing fraction of the universe bits represent information associated with the very high temperatures of stellar interiors, $>10^7$K. Note that the relative difference between the interior temperatures of stars of different types and at different nuclear burning stages, or ages, is small compared with the many orders of magnitude these stellar interior temperatures exceed the temperature of non-stellar matter, primarily in the form of gas and dust. Accordingly, during the early rapid stellar evolution the average temperature and hence average characteristic information bit energy changed primarily in proportion to the fraction of all bits accounting for information in stars. As this information accounts for the Boltzmann microstates of matter, we expect that fraction equals the fraction of all baryons that are in stars.

The upper plot in fig. 1 reproduces results from the recent (2006) survey of measurements of high redshift galaxy populations by Hu and Cowie [24]. The fraction of all baryons, $f_b$, that are in stars has been replotted as a function of universe scale size, $a$, (replotted from figure 1 of [24]). To simplify modelling the data has been split into two regions by the vertical line: the main region of rapid change with redshifts, $z>0.8$; and a more constant region $z<0.8$. Data points for $z>0.8$ are reasonably well described by a power law variation with the best fit for $f_b$ (solid line) corresponding to $a^{+3.6}$ and the best fit for the specific variation as $a^{+3.0}$ is also drawn (dashed line) for comparison. A detailed fit to the data might entail at least three separate periods of fit with the steepest period in the middle, but, for data $z>0.8$, it is sufficient for our purposes to show that $f_b$ varies close to $a^3$.

The lower plot in fig. 1 illustrates how this increasing fraction of baryons in stars affected the average baryon (and bit) temperature. The majority of baryons, non stellar gas and dust, continued to cool as $a^{-0.75}$ reaching a temperature today only one order of magnitude higher than the 2.7K temperature of the cosmic microwave background radiation, CMB, consistent with the present matter energy density four orders of magnitude higher than the CMB energy density (see equation 1). However, the proportion of baryons in stars increased as $a^{+3.0}$ reaching a maximum of around 10% at $z=0.8$, causing the average bit temperature to also increase as $a^{+3.0}$ reaching a temperature of a few times $10^6$K (10% of typical stellar temperatures) at z=0.8.

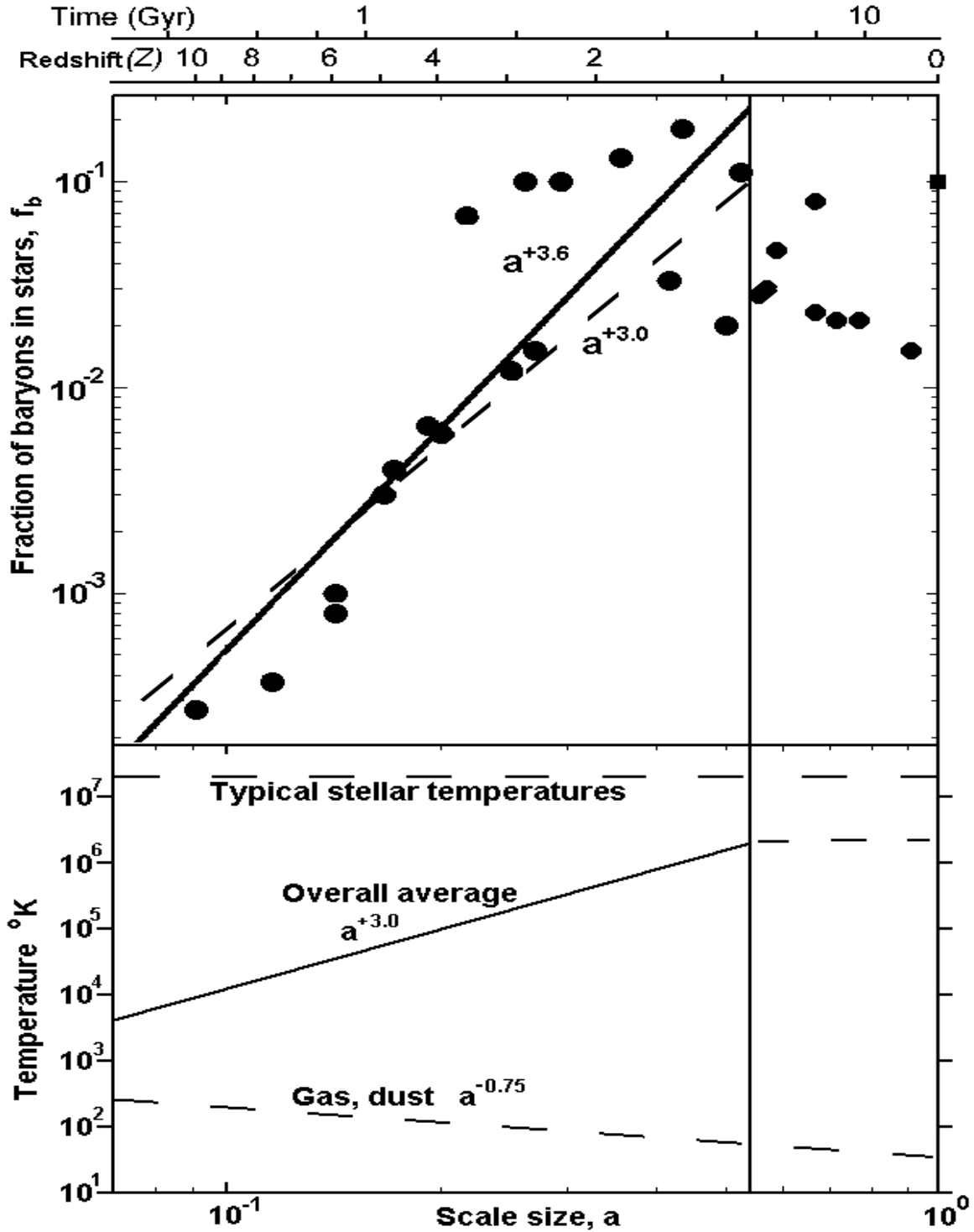

**Figure 1.** *Upper plot.* The cumulative fraction of all baryons, $f_b$, that have formed stars plotted as a function of universe scale size, $a$, from Hu & Cowie [24]. The black line is the best power law fit for data $z > 0.8$ (vertical line limit) and the dashed line the best $a^{+3.0}$ fit. The reader is referred to [24] for details of all experimental techniques and corrections applied in this survey of measurements.
*Lower plot.* The increasing fraction of baryons found in stars during the period $10 > z > 0.8$ causes the overall average baryon (and bit) temperature to also increase as $a^{+3.0}$.

An average bit temperature rising as $a^{+3.0}$ is the specific characteristic needed to counteract the $a^{-3.0}$ bit density dilution caused by the increasing separation between formed stars as the universe expands. Thus, for the period $10>z>0.8$, there was a near constant overall information energy density ($a^0$), or an information equation of state, $w_i \sim -1.0$.

In recent times $z<0.8$ the data points vary less and the fraction of baryons in stars has reached a maximum, $\sim 10^{-1}$, necessarily limited by definition to $f_b <1$. For reference, the black square in fig. 1 corresponds to the present day estimated fraction of all baryons that are in galaxies [24-25].

## 5. Overall time history of $w_i$

The overall time history of the information equation of state is summarised in fig. 2. Before star formation commenced the combination of expansion and cooling lead to an information equation of state lying between the limits: $-0.75<w_i<+0.25$, with a most likely value of $w_i = +0.25$. During stellar formation the proportion of high temperature bits increased close to $a^{+3}$ counteracting the $a^{-3}$ dilution effects from increasing star separations as the universe expanded. This resulted in a nearly constant overall information energy density, or $w_i \sim -1.0$, for the major part of cosmic time, $10>z>0.8$. During this rapid growth in stars, the subsequent changes in stellar temperature with star type and age after birth would have had little effect on the average bit temperature compared with the large temperature increase at star formation.

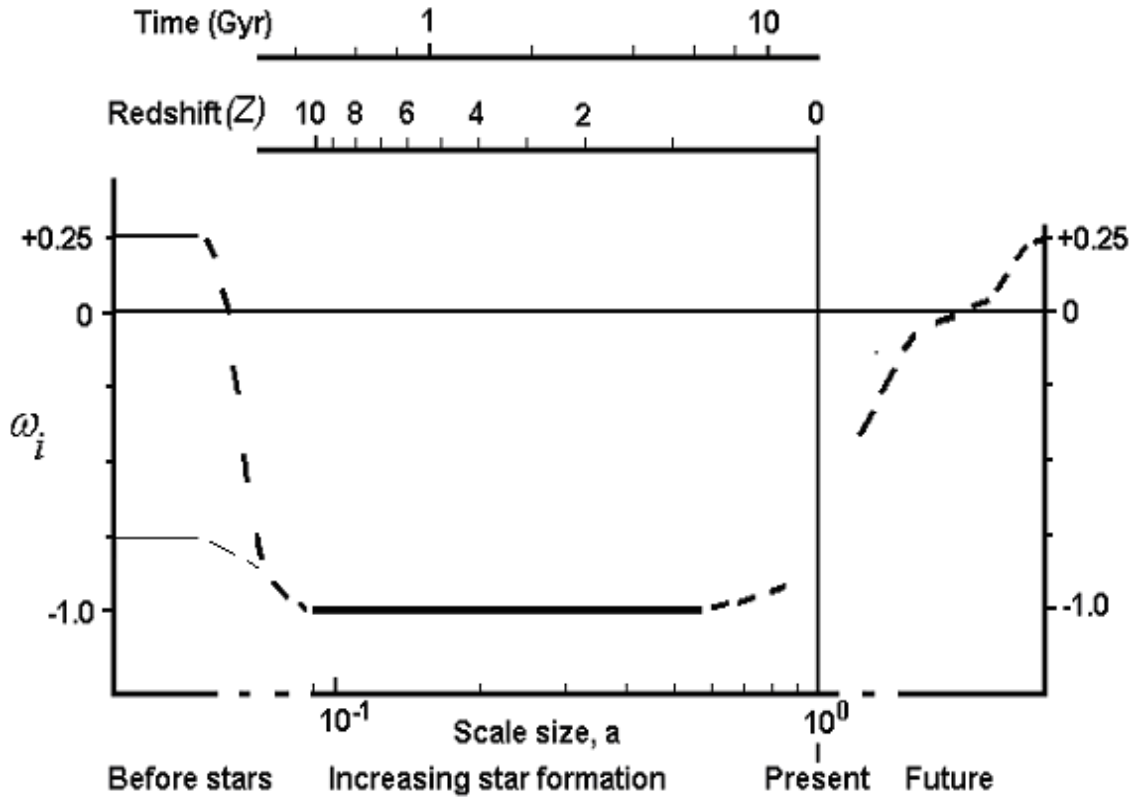

**Figure 2.** Time variation of the information equation of state, $w_i$.

For recent times, $z<0.8$, the previous rapid growth in $f_b$ slowed to reach a limit with the fraction of all baryons that are in stars $\sim 10^{-1}$ today as star birth and death rates become similar. Now the distribution of star types and temperature evolution with age becomes important. A detailed account of these effects to estimate the variation in average bit temperature in this recent period is beyond the scope of this work. We can assume that $w_i$ has remained significantly negative, at least initially, after $z=0.8$ as the average temperature continued to increase, primarily determined by the stars formed around $z\sim0.8$ as they move through hydrogen burning to hotter helium and later burning stages. In the future the overall distribution of stellar temperatures is expected to change less, leading to a more constant average temperature and a tendency for $w_i \rightarrow 0$.

## 6. Discussion

We have shown that the information equation of state was negative and equal to the dark energy value $w_{DE} \sim -1.0$ during the rapid star formation period $10>z>0.8$, or for over one half of cosmic time. Any negative equation of state, and specifically the value $w_i \sim -1.0$, implies that information energy must make a contribution to dark energy.

The universe information energy total is given by the product of total universe information content with the average information bit energy. The lower plot of fig. 1 shows that the average bit temperature since $z=0.8$ is around $2\times10^6$K (~10% of typical stellar temperatures) and thus the average bit energy around 120eV, from $k_B\,T\,ln2$. Then, in order for the information energy to account for all dark energy, the total universe information content would need to be $\sim 10^{87}$ bits. This value is not unreasonable and is similar to the previous estimates of $10^{90}$ bits [18][20]. However, care must be taken to avoid a circular argument and not to rely solely on this similarity to support our argument since the $10^{90}$ bits value of the previous work [18-20] was effectively estimated by dividing the universe total mass energy by the characteristic energy of our equation (2).

Any significant information energy contribution to dark energy is directly relevant to the cosmic coincidence problem as it implies a natural coupling between dark energy and matter. In particular, the information component of dark energy identified here is directly related to the degree of stellar evolution, consistent with dark energy only affecting the universe since sometime after stars began forming. In providing some response to the cosmic coincidence 'Why now?' question we are lead to the important 'What next?' question. A full calculation of the average bit temperature after $z=0.8$ requires a sum over all star types taking into account their different temperature histories and will be the subject of subsequent work. Nevertheless, we can expect $w_i \rightarrow 0$ in the future since the extent of stellar formation has reached a maximum. Whilst we expect dark energy to remain the dominant energy component of the universe for some time, dark energy density will no longer remain constant but will eventually fall along with mass energy density as $a^{-3}$, or $w_i=0$. The universe will continue to expand but no longer with an accelerating expansion.

In common with previous work [18-20] we have concentrated on information associated with baryons, ignoring the information energy density contributions from other universe components such as dark matter or the cosmic microwave background, CMB. Other components may have similar information bit contents but their information energy densities will not be significant as characteristic temperatures are orders of magnitude below stellar temperatures. Dark matter is thought to be present at a mass density ~5x the mass density of ordinary matter. Most explanations for dark matter involve either cold dark matter models or primarily cold models involving a mixture of mostly cold with some hot dark matter. In any case, the average dark matter temperature is thought to be $<<10^7$K. CMB can be considered a 'snapshot' of the early universe, effectively possessing an information content equal to that of the early universe, and therefore similar in value to today's universe, assuming CMB and the universe have both conserved information content in the intervening period. Then the very low 2.7K CMB temperature means the CMB information energy density is also insignificant compared with the hot stellar baryon component.

In the above estimation of $w_i$ during stellar evolution we have ignored the loss of entropy, or reduction of information, resulting from the reduction in the number of microstates assumed to take place on star formation as initially highly disordered matter collapses to form the more ordered structure of a star [26]. However, this reduction in bit number is small compared with the major increase in temperature at star formation and will have no effect on our calculated equation of state parameter. We can illustrate this by assuming the information contained by any physical system must be similar to the information required to fully simulate that system on a computer. For example, a full simulation of baryons during star formation would require a resolution at least of the order of the Planck length, $1.6 \times 10^{-35}$ m. Consider changing from describing a baryon location within the universe ($\sim 10^{26}$m) to a location within a typical star, for example the sun ($\sim 10^9$m). This corresponds to a change in accuracy from one part in $6\times 10^{60}$ ($\sim 2^{202}$) to one part in $6 \times 10^{43}$ ($\sim 2^{145}$). In order to maintain accuracy sufficient for a full simulation the minimum number of bits required to represent that location parameter on the computer only changes from 202 bits to 145 bits. Such a small 28% reduction in information is miniscule compared with a typical five orders of magnitude increase in temperature at star formation.

The reduction in thermodynamic entropy on star formation is usually made compatible with the second law and the 'arrow of time' by assuming there is at least a compensating increase in gravitational entropy [26]. Alternatively, Landauer's principle leads us to expect that the reduction of entropy, or information 'erasure' from reduction of Boltzmann microstates, on star formation dissipates heat to raise the entropy of the surrounding stellar environment in compensation and in accord with the second law of thermodynamics and conservation of information.

It has been noted before [27] that equation (2) above for the characteristic energy, $E$, of an information bit in the universe, excluding star formation, is identical in form and value to the characteristic energy of a cosmological constant, one of the possible descriptions for dark energy. Taking $ln2 \sim 1$, equation (2) is <u>identical</u> to equation 17:14 of reference [28]. With the present mass density equation (1) gives an average gas and dust temperature of 35K, and equation (2) provides the $3\times 10^{-3}$eV bit energy or characteristic energy that has been associated [28] with a cosmological constant. Thus our information based argument naturally explains this low value as $k_B\ T\ ln2$ – a value previously thought 'difficult to explain as it is too small to relate to any interesting particle physics' [28].

Finally, fig 1. clearly shows that the earlier rapid $a^3$ rise in the fraction of baryons in stars ceased around $z$=0.8. Although those stars formed at that time will continue to evolve through higher temperature nuclear cycles the average bit temperature clearly no longer rises as steeply as $a^3$. We therefore expect to see some observable change to the dark energy equation of state towards a value $w>-1$ in the most recent times. The contribution information energy makes to dark energy might therefore be verified experimentally by one of the future experiments planned to measure the effects of dark energy with greater precision.

## 7. Summary

By considering a simple model for the equation of state of information in the universe we have shown that information energy must make a significant contribution to dark energy. While the approach taken here may be considered unconventional, the conclusion is supported by a number of strong arguments:

- The information equation of state was clearly negative for at least one half of cosmic time, $10>z>0.8$, with a value close to the dark energy $w \sim -1.0$.
- Information energy can easily account for all of the dark energy with an information bit content of $\sim 10^{87}$ bits, similar in magnitude to previous estimates for the universe bit content.
- The equation for the characteristic bit energy of non-stellar information is identical in form to the equation for the characteristic energy associated with a cosmological constant.
- The low characteristic energy associated with dark energy was previously thought difficult to explain from particle considerations but is a natural result of this information approach.
- Information energy is directly related to the degree of stellar evolution, and thus can provide an answer to the fundamental cosmic coincidence question 'Why now?'
- Occam's razor argues strongly for such a simple explanation to the main (70%) energy component of the universe.
- Whether information energy is the source of dark energy can be tested experimentally by searching for evidence a dark energy value $w>-1.0$ in the most recent period ($z<<0.8$).

**Acknowledgements**

The author would like to acknowledge previous discussions with Prof. Ed Copeland (University of Nottingham), Dr. Tobia Carozzi (University of Glasgow), and Dr. Andrew Buckley (University of Sussex).

**References and Notes**

1   Landauer R, *IBM J. Res. Development*, **1961**, *3*, 183-191
2   Bennett C H, *IBM J. Res. Development*, **1973**, *17*, 525-532
3   Bennett C H, *Int. J. Theoretical Phys.,* **1982**, *21*, 905-940
4   Bennett C H, *IBM J. Res. Develop.,* **1988**, *32,* 16-23.
5   Landauer R, *Nature,* **1988,** *335,* 779-784.

6　Landauer R, *Physica Scipta*, **1987**, *35*, 88-95
7　Bennett C H, *Superlattices and Microstructures*, **1998**, *23*, 367-372
8　Feynman R P, *Lectures on Computation*, **1999**, 137-184, Penguin Books.
9　Bennett C H, *Studies in History and Philosophy of Modern Physics*, **2003**, *34*, 501-510
10　Shannon C E, *Mathematical Theory of Communication*, **1963**, U. Illinios Press.
11　Plenio M B and Vitelli, *Contemporary Physics*, **2001**, *42,* 25-60
12　Ladyman J et al, *Studies in History and Philosophy of Modern Physics*, **2007**, *38*, 58-79
13　Piechocinska B, *Phys. Rev,* **2000***, A 61*, 062314,1-9
14　Daffertshofer A and Plastino A R, *Physics letters A*, **2005**, *342*, 213-216
15　Zeilinger, A, *Found. of Phys.***, 1999**, *29*, 631-643
16　Braunstein S L and Pati A K, *Preprint gr-qc/*0603046., **2006,** 1-4
17　Lee J-W, Lee J, and Kim H.C, *Preprint hep-th/*07090047, **2007,** 1-4
18　Lloyd S, *Nature,* **2000**, *406*,1047-1054.
19　Lloyd S, *Programming the Universe,* **2006***,* Alfred A. Knopf.
20　Lloyd S, *Phys Rev Lett*., **2002**, *88*, 237901, 1-4
21　Bekenstein J D, *Phys. Rev., **1973**, D 7,* 2333-2346
22　Treumann RA, *Astrophys and Space Sci*.**,1993**, *201*, 135-147
23　Cirkovic M M, *Found. of Phys.*, **2002**, *32*, 1141-1157
24　Hu E M and Cowie L L, *Nature* **2006**, *440***,** 1145-1150.
25　Cole S,et al, *Mon. Not. R. Astron Soc* **2001**, *326***,** 255-273.
26　Penrose R, *The Road to Reality*, **2004**, 705-707, Jonathan Cape, London
27　Gough M P, Carozzi T, Buckley A M, *Preprint astro-ph*/0603084, **2006,** 1-6
28　Peebles P J E, *Principles of Physical Cosmology*, **1993**, 399, Princeton University Press.